\documentclass[twocolumn,letterpaper,10pt]{article}
\usepackage{iasted}
\usepackage{times}
\usepackage[dvips]{graphicx}
\usepackage{amssymb,amsmath}

\begin{document}

\date{}

\title{AN EXPLORATION OF CUDA AND CBEA FOR A GRAVITATIONAL WAVE SOURCE-MODELLING APPLICATION}

\author{
Gaurav Khanna \\
Physics Department\\
University of Massachusetts at Dartmouth\\
285 Old Westport Rd.\\
Dartmouth, MA 02747 \\
email: gkhanna@umassd.edu \\
\and
Justin McKennon\\
Electrical and Computer Engineering Department\\
University of Massachusetts at Dartmouth\\
285 Old Westport Rd.\\
Dartmouth, MA 02747 \\
email: jmckennon@umassd.edu \\
}

\maketitle

\thispagestyle{empty}

\noindent
{\bf\normalsize ABSTRACT}\newline
{In this paper, we accelerate a gravitational physics numerical modelling application using hardware accelerators -- Cell processor and Tesla CUDA GPU. We describe these new technologies and our approach in detail, and then present our final performance results. We obtain well over an {\it order-of-magnitude} performance gain in our application by making use of these {\em many}-core architectures.} \vspace{2ex}
   
\noindent
{\bf\normalsize KEY WORDS}\newline
{CUDA, GPU, Cell, CBEA, modelling, gravitation}

\section{Introduction}

Computational scientists have recently begun making use of hardware accelerators (such as the Cell processor and CUDA GPUs) because these can provide significant gains in the overall performance of many numerical simulations at a relatively low cost. Cell Broadband Engine Architecture (CBEA) [1] was designed by collaboration between Sony, Toshiba and IBM and is being used in Sony's PlayStation 3 (PS3) [2] and high-performance computing hardware (LANL RoadRunner [3]). Compute Unified Device Architecture (CUDA) [4] is Nvidia's general-purpose software development system for their current GPUs. 

In this work, we make use of these new technologies to accelerate an application from the numerical relativity (NR) community -- the EMRI Teukolsky code [9] which is a finite-difference, linear, hyperbolic, inhomogeneous partial difference equation (PDE) solver. In this paper, we illustrate in detail our implementation and approach to utilizing the low-level  data parallelism that is offered by these architectures.

It is worth pointing out that our NR application is of a type that is quite common in various fields of science and engineering, therefore we expect that our work would be of interest to the larger community of computational scientists. These architectures have been recently evaluated for other numerically intensive problems, and their performances have been compared and presented in the relevant literature [10 -- 13]~\footnote{It should be noted that Ref. [13] is about a similar exploration of CBEA and CUDA for a gravitational wave data-analysis code, which is a completely different application compared to the one under consideration here. The data-analysis application in Ref. [13] searches for gravitational waveforms present in real observational data; while the source-modelling application that we are describing in this article, is one that uses a theoretical model of a gravitational wave source to generate numerical waveform data.}.

This paper is organized as follows: In Section 2, we introduce the EMRI Teukolsky code, the relevant background gravitational physics and the numerical method used by the code.  It is this code that we accelerate in our work using hardware accelerators. In Section 3, we provide a general introduction to the Cell processor and CUDA GPUs. We emphasize aspects of these hardware accelerators that are relevant to our implementations, which are detailed in Section 4. In Section 5, we present the overall performance results from our parallelization efforts. Finally, in Section 6, we summarize this work and make some conclusive remarks.
 
\section{EMRI Teukolsky Code}

Many gravitational wave observatories [9] are currently being built all over the globe. These laboratories will open a new window into the Universe by enabling scientists to make astronomical observations using a completely new medium -- gravitational waves (GWs) as opposed to electromagnetic waves (light). These GWs were predicted by Einstein's relativity theory, but have not been directly observed because the required experimental accuracy was simply not advanced enough (until very recently). 

Numerical relativity is an area of gravitational physics that is focussed on the numerical modelling of strong sources of GWs -- collisions of compact astrophysical objects such as neutron stars and black holes. Thus, it plays an extremely important role in this new and upcoming area of GW astronomy. The specific NR application that we consider in this paper is one that evolves the GWs generated by a compact object (such as a star of the size of our own Sun) that has a decaying orbit around a supermassive black hole. Such large black holes -- often more massive than a million times our Sun -- lurk at the center of most galaxies and routinely devour smaller stars and black holes. Such processes are commonly referred to as extreme mass-ratio inspirals (EMRIs) in the relevant literature. The low-frequency gravitational waves emitted from such EMRI systems are expected to be in good sensitivity band for the upcoming space-borne gravitational wave detectors -- such as the ESA/NASA Laser Interferometer Space Antenna (LISA) mission [6]. Studies of the dynamics and the orbital evolution of a binary system in the extreme mass-ratio limit is therefore an important issue for low-frequency gravitational wave detection. 

Because of the extreme mass-ratio, the small object orbiting around the central supermassive black hole can be modeled as a small structure-less object, and the problem can be addressed within black hole perturbation theory. This is where the Teukolsky equation becomes relevant. This equation governs the evolution of the perturbations of rotating (Kerr) black holes, with the small object acting as a ``source'' of the perturbations. In other words, the Teukolsky equation is essentially a linear wave equation in Kerr space-time geometry, with the small object acting as generator of the gravitational waves. Thus, to numerically model an EMRI scenario, we solve the inhomogeneous Teukolsky equation in the time-domain.

The next two subsections provide more detailed information on this equation and the associated numerical solver code.

\subsection{Teukolsky Equation}

The Teukolsky master equation describes scalar, vector and tensor field perturbations in the space-time of Kerr black holes [14]. In Boyer-Lindquist coordinates, this equation takes the form
\begin{eqnarray}
\label{teuk0}
&&
-\left[\frac{(r^2 + a^2)^2 }{\Delta}-a^2\sin^2\theta\right]
         \partial_{tt}\Psi
-\frac{4 M a r}{\Delta}
         \partial_{t\phi}\Psi \nonumber \\
&&- 2s\left[r-\frac{M(r^2-a^2)}{\Delta}+ia\cos\theta\right]
         \partial_t\Psi\nonumber\\  
&&
+\,\Delta^{-s}\partial_r\left(\Delta^{s+1}\partial_r\Psi\right)
+\frac{1}{\sin\theta}\partial_\theta
\left(\sin\theta\partial_\theta\Psi\right)+\nonumber\\
&& \left[\frac{1}{\sin^2\theta}-\frac{a^2}{\Delta}\right] 
\partial_{\phi\phi}\Psi +\, 2s \left[\frac{a (r-M)}{\Delta} 
+ \frac{i \cos\theta}{\sin^2\theta}\right] \partial_\phi\Psi  \nonumber\\
&&- \left(s^2 \cot^2\theta - s \right) \Psi = -4\pi (r^2 + a^2 \cos^2 \theta)\, T ,
\end{eqnarray}
where $M$ is the mass of the black hole, $a$ its angular momentum per unit mass, $\Delta = r^2 - 2 M r + a^2$ and $s$ is the ``spin weight'' of the field. The $s = \pm 2$ versions of these equations describe the radiative degrees of freedom of the gravitational field, and thus are the equations of interest here. As mentioned previously, this equation is an example of linear, hyperbolic, inhomogeneous PDEs that are quite common in several areas of science and engineering, and can be solved numerically using a variety of finite-difference schemes. The quantity $T$ in Eq. (1) is the ``source'' term as mentioned in the previous section. It plays an extremely critical role in this work and that will be discussed in detail, later in this paper. Ref. [14] has a mathematical formula for this quantity and to save space, we will not reproduce that expression here.

\subsection{Numerical Method}

Ref. [15] demonstrated stable numerical evolution of Eq.\ (\ref{teuk0}) for $s=-2$ using the well-known Lax-Wendroff numerical evolution scheme. Our Teukolsky code uses the exact same approach, therefore the contents of this section are largely a review of the work presented in the relevant literature [9]. 

Our code uses the tortoise coordinate $r^*$ in the radial direction and azimuthal coordinate $\tilde{\phi}$. These coordinates are related to the usual Boyer-Lindquist coordinates by
\begin{eqnarray}
dr^* &=& \frac{r^2+a^2}{\Delta}dr 
\end{eqnarray}
and
\begin{eqnarray}
d\tilde{\phi} &=& d\phi + \frac{a}{\Delta}dr \; . 
\end{eqnarray}  
Following Ref. [9], we factor out the azimuthal dependence and use the ansatz,
\begin{eqnarray}
\label{eq:psiphi}
\Psi(t,r^*,\theta,\tilde{\phi}) &=& e^{im\tilde{\phi}} r^3 \Phi(t,r^*,\theta) .
\end{eqnarray}
Defining
\begin{eqnarray}
\Pi &\equiv& \partial_t{\Phi} + b \, \partial_{r^*}\Phi \; , \\
b & \equiv &
\frac { {r}^{2}+{a}^{2}}
      { \Sigma} \; , 
\end{eqnarray}
and
\begin{eqnarray}
\Sigma^2 &\equiv &  (r^2+a^2)^2-a^2\,\Delta\,\sin^2\theta
\; 
\label{pi_eq}
\end{eqnarray} 
allows the Teukolsky equation to be rewritten as
\begin{eqnarray}
\label{eq:evln}
\partial_t \mbox{\boldmath{$u$}} + \mbox{\boldmath{$M$}} \partial_{r*}\mbox{\boldmath{$u$}} 
+ \mbox{\boldmath{$Lu$}} + \mbox{\boldmath{$Au$}} =  \mbox{\boldmath{$T$}} ,
\end{eqnarray}
where 
\begin{equation}
\mbox{\boldmath{$u$}}\equiv\{\Phi_R,\Phi_I,\Pi_R,\Pi_I\}
\end{equation}
is the solution vector. The subscripts $R$ and $I$ refer to the real and imaginary parts respectively (note that the Teukolsky function $\Psi$ is a complex valued quantity). Explicit forms for the matrices {\boldmath{$M$}}, {\boldmath{$A$}} and {\boldmath{$L$}} can be easily found in the relevant literature [9]. Rewriting Eq.\ (\ref{eq:evln}) as 
\begin{equation}
\partial_t \mbox{\boldmath{$u$}} + \mbox{\boldmath{$D$}}
\partial_{r^*} \mbox{\boldmath{$u$}}
=  \mbox{\boldmath{$S$}}\; , 
\label{new_teu2}
\end{equation}
where
\begin{equation}
 \mbox{\boldmath{$D$}} \equiv \left(\begin{matrix}
                    b &   0   &  0  &  0 \cr
                    0  &   b   &  0  &  0 \cr
                    0  &   0   &  -b  &  0 \cr
                    0  &   0   &  0  &  -b \cr
                \end{matrix}\right),
\label{d_matrix}
\end{equation}
\begin{equation}
\mbox{\boldmath{$S$}} =\mbox{\boldmath{$T$}} -(\mbox{\boldmath{$M$}} - \mbox{\boldmath{$D$}})
\partial_{r^*}\mbox{\boldmath{$u$}}
- \mbox{\boldmath{$L$}}\mbox{\boldmath{$u$}} 
- \mbox{\boldmath{$A$}}\mbox{\boldmath{$u$}},
\end{equation}
and using the Lax-Wendroff iterative scheme, we obtain stable evolutions. Each iteration consists of two steps: In the first step, the solution vector between grid points is obtained from
\begin{eqnarray}
\label{lw1}
\mbox{\boldmath{$u$}}^{n+1/2}_{i+1/2} &=& 
\frac{1}{2} \left( \mbox{\boldmath{$u$}}^{n}_{i+1}
                  +\mbox{\boldmath{$u$}}^{n}_{i}\right)
- \\
&  &\frac{\delta t}{2}\,\left[\frac{1}{\delta r^*} \mbox{\boldmath{$D$}}^{n}_{i+1/2}
  \left(\mbox{\boldmath{$u$}}^{n}_{i+1}
                  -\mbox{\boldmath{$u$}}^{n}_{i}\right)
- \mbox{\boldmath{$S$}}^{n}_{i+1/2} \right] \; .\nonumber
\end{eqnarray}
This is used to compute the solution vector at the next time step,
\begin{equation}
\mbox{\boldmath{$u$}}^{n+1}_{i} = 
\mbox{\boldmath{$u$}}^{n}_{i}
- \delta t\, \left[\frac{1}{\delta r^*} \mbox{\boldmath{$D$}}^{n+1/2}_{i}
  \left(\mbox{\boldmath{$u$}}^{n+1/2}_{i+1/2}
                  -\mbox{\boldmath{$u$}}^{n+1/2}_{i-1/2}\right)
- \mbox{\boldmath{$S$}}^{n+1/2}_{i} \right] \, .
\label{lw2}
\end{equation}
The angular subscripts are dropped in the above equation for clarity. All angular derivatives are computed using second-order, centered finite difference expressions. 

Following Ref. [9], we set $\Phi$ and $\Pi$ to zero on the inner and outer radial boundaries. Symmetries of the spheroidal harmonics are used to determine the angular boundary conditions: For even $|m|$ modes, we have $\partial_\theta\Phi =0$ at $\theta = 0,\pi$ while $\Phi =0$ at $\theta = 0,\pi$ for modes of odd $|m|$.
 
\begin{figure}
\centering
\includegraphics[width=3in]{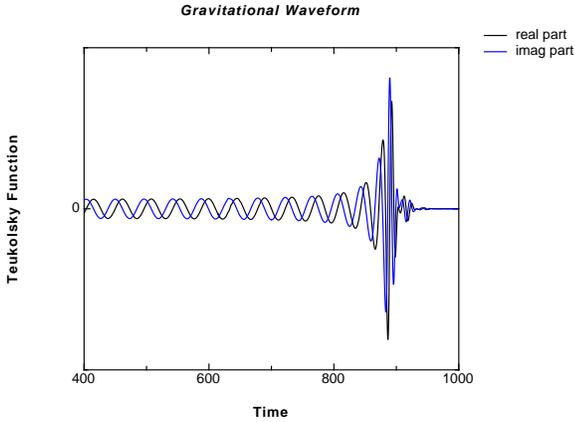}
\caption{Sample gravitational waveform ($\Phi$) results from the EMRI Teukolsky code.}\label{evol}
\end{figure}

As a sample numerical result, we show the gravitational wave signal from a binary of mass-ratio $10^{-4}$ with a decaying circular-equatorial orbit. The central black hole has zero angular momentum (spin). This smaller member of the binary, slowly begins to spiral in and then at a much later time, plunges into the central black hole causing it to release a burst of gravitational radiation. Fig.~\ref{evol} shows the results, illustrating the $m = 2$ mode of the gravitational waveform.

\section{Cell/GPU Hardware Accelerators}

Computer simulations are playing an increasingly important role in science and engineering today. The driving force behind this fact is the ability to reproduce costly laboratory experiments with low cost computer hardware based simulations. The reason why numerically intensive simulations are relatively very inexpensive is because of the rapid increases in the overall performance ({\it Moore's Law}) of mass-produced computer hardware over the past several decades. 

However, a few years ago the entire computer hardware industry hit a serious {\it frequency wall}, implying that increasing a processor's clock-rate for gains in performance could not be done indefinitely, due to rapid increases in power consumption and heat generation ({\it power wall}). This led all the major processor manufactures toward multi-core designs. Today all commodity laptop and desktop processors are multi-core i.e. they combine two or more independent computing cores on a single die. Thus, manufacturers are now able to pack even more power into a processor without having to increase its clock frequency.

Now, it turns out that there are other computing technologies that are based on a {\it many}-core design, and their overall performance has continued to increase at a rate much higher than that of traditional processors. These technologies have typically been employed in desktop graphics cards and consumer gaming consoles. In this section we will briefly describe these technologies -- STI Cell Broadband Engine (CBE) [1] and Nvidia Tesla CUDA GPU [4] -- with emphasis on specific aspects that are relevant to our EMRI Teukolsky code's development.  

\subsection{Cell Processor}

The CBE [1] is a totally redesigned processor that was developed collaboratively by Sony, IBM and Toshiba primarily for multimedia applications. This processor has a general purpose (PowerPC) CPU, called the PPE (that can run two (2) software threads simultaneously) and eight (8) special-purpose compute engines, called SPEs available for raw numerical computation. Each SPE can perform vector operations, which implies that it can compute on multiple data, in a single instruction (SIMD). All these compute elements are connected to one another through a high-speed interconnect bus (EIB). Note that the design of this processor is very different from traditional multi-core processors. In a certain sense, the CBE's design is somewhere between a general-purpose CPU and a specialized GPU (as described in the next subsection). It can therefore be considered as {\it hybrid} technology, having the advantages of both these architectures. The outcome of this distinctive design is that a single, 3.2 GHz (original -- 2006/2007) CBE has a peak performance of over 200 GFLOP/s in single-precision floating-point computation and 15 GFLOP/s in double-precision operations. It should be noted that the current (2008) release of the CBE, called the PowerXCell, has design improvements that bring the double-precision performance up to 100 GFLOP/s.

We will not attempt to go into much more detail concerning the CBE's design here, rather we will simply point out one unique feature that addresses the issue of the {\it memory wall} that is common to all current computer hardware. The memory wall refers to the large (and increasing) gap between processor and memory performance, causing the slower memory speeds to become a significant bottleneck. The current state-of-the-art approach to combat this issue has been to include large cache sizes (several MBs) on the processor chip. However, this takes away valuable space (for compute elements) on the processor die, and thus may result in only a marginal, overall increase in performance. A key feature of the CBE is its unique ability to {\it interleave} computation and data access. Therefore, it is possible for the programmer to overlap memory access and the actual computation -- so-called ``double buffering'' -- in order to hide the time it takes to access memory. It is this mechanism that allows the CBE to break through the memory wall and perform very efficiently, even for computations that have a large memory footprint. It is also partly for this reason, that the CBE can reach a ``real world'' application performance that is nearly 100\% of its theoretical peak performance [7]. 

One challenge introduced by this new design, is that the programmer has to explicitly manage the data transfer between the PPE and the SPEs. The PPE and SPEs are equipped with a DMA engine -- a mechanism that enables data transfer to and from main memory and each other. Now, the PPE can access main memory directly, but the SPEs can only directly access their own rather limited (256KB) local store. Therefore, the SPE local store has to contain both the instructions and the data to compute on, at any given moment. In our implementation, we load the instructions onto the SPEs during the initialization stage once and for all, and then we move data through as needed, during the numerical evolution. 

Another important mechanism that allows communication between the different elements (PPE, SPEs) of the CBE is the use of mailboxes. These are special purpose registers that can be used for uni-directional communication. Each SPE has three (3) mailboxes -- two (2) outbound, that can hold only a single entry, and one (1) inbound, that can hold four (4) entries. These are typically used for synchronizing the computation across the SPEs and the PPE, and that is primarily how we make use of these registers as well.

The parallel programming model on CBEA allows for the use of SPEs for performing different tasks in a workflow (``task parallel'' model) or performing the same task on different data (``data parallel'' model). We make use of the data parallel model in our EMRI Teukolsky code implementation.

\subsection{CUDA GPUs}

Nvidia's CUDA [4] is a set of software layers that communicate with its GPUs -- at a high level: CUDA Runtime API, and at a low level: CUDA Driver API. Since the high level API is implemented above the lower level, each function call of the Runtime API is broken into simpler instructions and managed by the Driver API.

Through CUDA, the GPU (called {\it device}) is accessible to the CPU (called {\it host}) as a co-processor with its own memory. The device executes a function (usually referred to as a {\it kernel}) in a data parallel model, which means that a number of {\it threads} run the same program on different data. A {\it warp} is a group of 32 threads, which is the smallest grouping of threads that is processed by a GPU. In CUDA, work is performed with {\it blocks} of 64 -- 512 threads that are assembled into grids. The number of blocks processed simultaneously is closely related to their grouping. The grid-like architecture of the GPU makes it possible to apply a kernel to a large quantity of threads in one single call without worrying about fixed resources. CUDA Runtime manages all of these resources for the programmer. If the hardware has a large number of processing units, it can process them in parallel. Likewise, if the hardware has fewer resources, the blocks will be executed sequentially. This idea means that the same code can work for entry level GPUs, high-level GPUs, and even future GPUs without any modifications to the code itself. In the area of high performance computing, this idea of massive parallelism is extremely important. In the event of many of the same calculations being done at the same time, the CUDA/GPU architecture allows one to farm all of these calculations into a series of blocks and execute them all in parallel. The threads then scatter, doing all of their farmed computations, and synchronize at the completion of their assigned tasks. 

The GPU provides significant flexibility in terms of memory management. Six (6) main types of memory exist in the form of {\it registers}, {\it local} memory, {\it shared} memory, {\it global} memory, {\it constant} memory and {\it texture} memory.

Registers are located on the GPU chip itself. They provide simple read/write capabilities and have no cache. The registers exist as the fastest form of memory on a multi processor and should be utilized whenever possible. They are only accessible by the thread and exist only during the lifetime
of the thread. Local memory is an abstraction of memory that doesn't actually exist. It simply implies that the memory being used exists only in the scope of a single thread. This type of memory exists in the global memory allocated by the compiler and performs at the same speed as normal global memory. Local memory tends to be one of the slowest forms of memory (up to 150x slower than register or shared memory) and is only accessible by the thread. When the lifetime of the thread is complete, local memory is reallocated into the global memory space. 

Shared memory exists as a read/write type of memory. All of the threads have access to shared memory which allows for the communication between threads. When no bank conflicts exist (multiple threads attempting to access the same pieces of data from the same memory location) shared memory performs equally as fast as register based memory. Shared memory exists solely during the life of the block and is freed upon the blocks termination.

Global memory is the single most important type of memory to a GPU user and CUDA programmer. This type of memory provides the highest memory bandwidth only in the event that global memory accesses are able to be coalesced within half of a warp. This allows the hardware to fetch and store data in the fewest number of transactions. If the transactions cannot be coalesced, separate memory transactions will be executed for each thread in the half warp that inhibits efficient program execution by up to 150x. Global memory is accessible from either the host or device, and exists throughout the lifetime of the application. Using this type of memory exclusively is highly undesirable due to the increased possibility of uncoalesced memory read/write instructions. The high cost of accessing global memory is again related to the memory wall that was mentioned previously (Section 3.1). However, in CUDA there is a different way to overcome this problem. The device uses an efficient thread scheduler that uses the massive parallelism of the device to hide the latency by removing threads that just issued a global memory read from its processor and scheduling a thread that is not waiting for data. This is one of the reasons why the device requires more threads than there are processors available to achieve good performance.

Constant memory is only read from kernels, and does not offer the ability to write to this type of memory. It is fully accessible by all threads and the host device itself. This type of memory is set up by the hardware in the event that all or multiple threads read from the same location. It provides only one cycle of latency when there is a cache hit despite the fact that it resides in device memory. If threads read from multiple locations, the accesses are serialized. Data access in constant memory can take anywhere from one cycle to hundreds depending on the locality of the cache.

Texture memory is provided by the processor to help accelerate frequently performed tasks. It is in this type of memory that the greatest performance gains can be realized. It is an alternative memory access path that programmers have the ability to bind to certain regions of GPU device memory. References to this type of memory can be bound to same, different, or overlapping textures in memory. Memory coalescence is not important in the usage of texture memory. Optimal performance can be achieved only when all of the threads inside of a warp read from locations that are spatially near one another.  Texture memory performs a read operation from global memory in the event of a cache miss. This makes it possible to theoretically exceed the memory bandwidth of the global memory through repeated use of the cache of the texture memory.

The Tesla GPU utilized in our calculations is designed with a massively parallel {\em many}-core architecture. The GPU itself has roughly 240 cores which potentially eliminates the need for a large computer cluster to solve complex problems. It features four (4) GBs of high speed memory. This allows large data sets to be stored locally for each processor and minimized data movement and transfers throughout the system. CUDA is supported on this device, which allows for the programming and design of applications to exploit the parallelism offered by the GPU. It adheres to the IEEE single- and double- precision floating-point units allowing for some of the highest and most precise floating-point calculations available on any one device. The Tesla utilizes shared data memory allowing groups of processor cores to communicate by means of low latency memory. Each processor core functions at 1.3 GHz and provides a standard memory bandwidth of approximately 100 GB/s.

\section{Cell/GPU Implementation}

The 8 SPEs of the CBE and the 240 cores of the Tesla GPU are the main compute engines of these devices, therefore one would want these to execute the most compute intensive tasks of a code in a data-parallel fashion. 

The first task in our work is to isolate the most compute intensive portions of our EMRI Teukolsky code. Upon performing a basic profiling of our code using the GNU profiler {\bf gprof}, we learn that simply computing the source-term $T$ (see Section 2.1) takes {\bf 99\%} of the application's overall runtime. Thus, it is natural to consider accelerating this $T$ calculation using data parallelization on the SPEs of the CBE and the cores of the Tesla GPU. We anticipate that this observation is fairly typical for codes of this broad type. 

A data-parallel model is straightforward to implement in a code like ours. We simply perform a domain-decomposition of our finite-difference numerical grid and allocate the different parts of the grid to different SPEs or GPU cores. On the Cell, each SPE computes the source-term $T$ for a single $r^*$ grid value, and the entire range of $\theta$ grid values. On the Tesla GPU, each thread computes $T$ for a single pair of $r^*$ and $\theta$ grid values. Note that all these calculations are independent, i.e. no communication is necessary between the SPEs/GPU threads. 

In addition, it is necessary to establish the appropriate data communication between the SPEs/GPU cores and the remaining code that is executing on the PPE/CPU respectively. On the CBE, we use DMA instructions (as mentioned before in Section 3.1) to move data between the SPEs and the PPE. On the Tesla CUDA GPU, we use {\bf cudaMemcpy} instructions to achieve the same. In total, only a rather small amount of data is required to be transferred back and forth from the SPEs/GPU at every time-step of the numerical evolution. To be more specific, approximately 10 floating-point numbers are required by the SPEs/GPU to begin the computation for a specific $(r^{*},\theta)$ and then they release only 2 floating-point values as the result of the source-term $T$ computation. Because of the rather modest amount of data-transfer involved, we do not make use of any advanced memory management features on both architectures. In particular, we do not make use ``double buffering'' on the CBE, and we only use global memory on the GPU. 

The source-term computation in itself is rather complicated -- the explicit mathematical expression for $T$ is too long to list here. A compact expression of its form can be found in Ref. [14] although that is perhaps of limited usefulness from the point of view of judging its computational complexity. It will suffice here to say that it is essentially a very long mathematical formula that is implemented using numerical code which uses elementary floating-point operations and approximately 4000 temporary variables of the complex data-type. The code makes use of {\it double-precision} floating-point accuracy because that is the common practice in the NR community and also a necessity for such finite-difference based evolutions, especially if a large number of time-steps are involved.  

In summary, it is worth pointing out that this Cell/CUDA implementation of our EMRI Teukolsky code is fairly straightforward. It should also be mentioned that we do not attempt to hand-tune the codes to tailor them for each architecture, in order to obtain maximal performance. Instead, we rely on the mature compiler suites to perform all low-level optimizations (such as vectorization on the SPEs) automatically. The CBE code development effort is somewhat more involved, and is therefore more time-consuming when compared to the CUDA effort. In the future, {\it OpenCL} [8] will perhaps make the process of porting and optimizing code on different hardware architectures and devices considerably easier. 

\section{Performance Results}

In this section of this paper, we report on the final results from our implementations, as outlined in the previous section (Section 4). We begin with the performance results from the CBE implementation, followed by the results from the CUDA case. Then, we report on the relative performance across all the architectures we have discussed in this paper -- CPU, CBE and CUDA GPU. 

We use the following hardware for our performance tests: IBM QS20 and QS22 blade systems, that sport the original Cell and the PowerXCell, respectively -- both clocked at 3.2 GHz. These systems are equipped with two (2) GBs main memory. For the CUDA case, our system sports the Nvidia C1060 Tesla CUDA GPU. This system has an AMD 2.5 GHz Phenom (9850 quad-core) processor as its main CPU and four (4) GBs of memory. All these systems are running Fedora Linux as the primary operating system. Standard open-source GCC compiler suite for code development is available on all these systems.

\subsection{Cell EMRI Teukolsky Code Performance}

\begin{figure}
\centering
\includegraphics[width=3in]{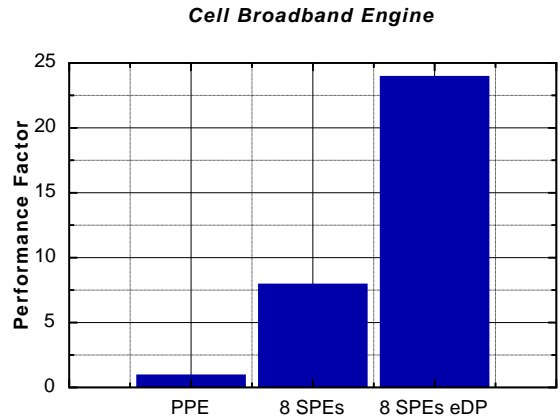}
\caption{Overall performance of the EMRI Teukolsky code accelerated by the Cell Broadband Engine. The baseline here is the CBE's PPE.}\label{cell}
\end{figure} 

In Fig.~\ref{cell} we show the overall performance results from our EMRI Teukolsky code as accelerated by the STI Cell processor. We choose the PPE as the baseline for this comparison. On the original (2006/2007) CBE, which has a rather limited capability in double-precision, we obtain an eight (8) fold gain in performance over the PPE. The current (2008) PowerXCell (labelled as eDP in the figure) that has the enhanced double-precision performance, delivers an impressive near 24 fold gain. Thus, by making use of the CBE hardware accelerator, we obtain an {\it order-of-magnitude} gain in our EMRI Teukolsky code's overall performance.   

In addition, it is worth noting that since we obtain very strong gains even from the the original release of the Cell processor -- the one found in the Sony PS3 -- one can obtain such performance gains using extremely low-cost hardware.

\subsection{CUDA EMRI Teukolsky Code Performance}

\begin{figure}
\centering
\includegraphics[width=3in]{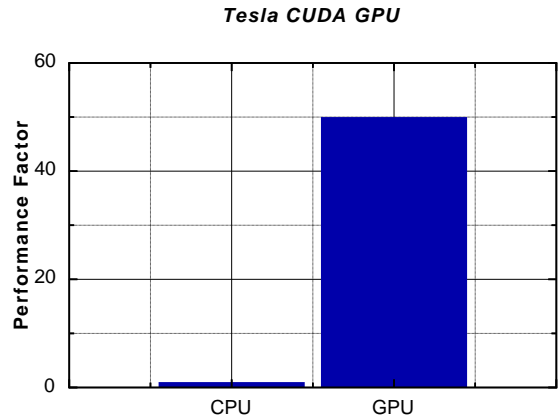}
\caption{Overall performance of the EMRI Teukolsky code accelerated by the Tesla CUDA GPU. The baseline here is the supporting system's CPU -- an AMD Phenom 2.5 GHz processor.}\label{gpu}
\end{figure} 

In Fig.~\ref{gpu} we show the overall performance results from our EMRI Teukolsky code as accelerated by the Nvidia Tesla CUDA GPU. Here we choose the CPU of the supporting system as the baseline. This CPU is a four (4) core AMD Phenom 2.5 GHz processor. We are unable to use all the cores of this multi-core CPU effectively. The CPU version of our code is OpenMP parallelized appropriately, but oddly enough that yields no benefit on this specific system (unlike on other multi-core systems) when compared with the plain serial version of the code. Therefore, we choose the baseline for these comparisons to be the best i.e. single-core (serial) performance of our EMRI Teukolsky code on an AMD 2.5 GHz Phenom processor.

As can be seen from the figure, our CUDA GPU acceleration results are very strong as well. In the double-precision context, which is our preferred level of accuracy, using the Tesla GPU enables a fifty (50) fold gain in the overall performance of the application. CUDA GPUs perform much better (by over an {\it order-of-magnitude}) on single-precision floating-point operations. In the near future, we expect the double-precision performance of GPUs to match their single-precision performance, and that hardware would deliver an extremely impressive acceleration of our application.     

To summarize our CUDA GPU results -- our implementation yields well over an {\it order-of-magnitude} gain in the overall performance of our EMRI Teukolsky code. In the future, as the double-precision performance of CUDA GPUs improves, we expect to obtain much stronger gains in our code's overall performance. 

\subsection{Relative Performance}

\begin{figure}
\centering
\includegraphics[width=3in]{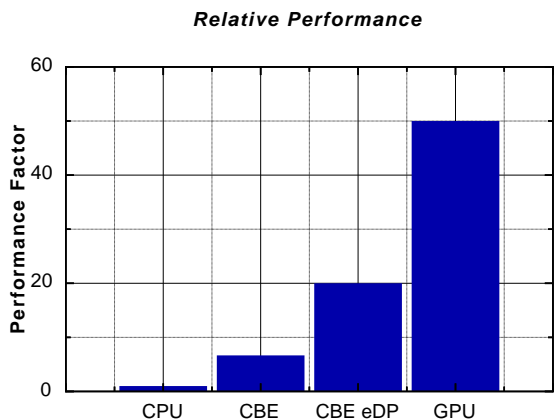}
\caption{Relative performance of the EMRI Teukolsky code on all discussed architectures -- CPU, CBE and GPU. The baseline here is the system CPU -- an AMD Phenom 2.5 GHz processor.}\label{comp}
\end{figure} 

In Fig.~\ref{comp} we depict the relative performance of all these architectures together, with the single-core 2.5 GHz AMD Phenom processor as the baseline. The results presented there are self explanatory. It is interesting to note that even if we were able to use all the four (4) cores of the Phenom processor extremely effectively, it would still be outperformed by all the hardware accelerators considered here, by a large factor. 

In addition to the statements about relative performance above, it is worth commenting on the comparative cost associated to procuring these different hardware architectures. The 2.5 GHz AMD Phenom is very inexpensive, and a system similar to the one we used for testing can be easily obtained for under \$1,000. The PowerXCell, that exhibits very impressive performance in our tests is the most expensive hardware to obtain, with its listed price on IBM's website being \$5,000 per processor. However, IBM is currently heavily discounting QS22 blades, and thus the ``street'' price is closer to \$2,500 per processor. It is worth pointing out that the original CBE, which also exhibits strong performance on our tests is available for a deeply discounted price of \$300 in the Sony PlayStation 3 game console [2]. Finally, the Nvidia Tesla GPU which exhibits the best performance on our tests, is currently available for approximately \$1,500. 

\section{Conclusions}

In this work, we take an important NR application -- the EMRI Teukolsky Code -- and perform a low-level parallelization of its most computationally intensive part, for optimized execution on the CBE and Tesla CUDA GPU. We describe the parallelization approach taken and also the relevant important aspects of the hardware accelerators in detail. 

The final outcome of our efforts is very similar on these two architectures -- we obtain well over an {\it order-of-magnitude} gain in overall application performance. However, it is worth mentioning that CBE/GPUs should not be considered general-purpose compute accelerators -- the actual computational problem and the relevant parallel approach has to be one that is well suited for the hardware's specific design, else one may gain very little. In general, the more massively parallel the task is, the better a design like that of a GPU would perform. For highly serial tasks, a traditional CPU is still the best option. And finally, for a task that is somewhere in between these two extremes, a design like that of the CBE may be optimal. From our viewpoint, because the CBE incorporates aspects of both GPU and CPU designs, it is very versatile and perhaps likely to perform very well on many different types of tasks. 

In the near future, the next generation of these hardware accelerators will be released -- the CBE will have more SPEs and a more powerful PPE; and CUDA GPUs will have much stronger double-precision performance and even more compute cores. Other processor manufacturers are also positioned to soon release similar hardware designs -- for example, Intel's Larrabee GPU. The overall performance of our EMRI Teukolsky code will benefit further from these improvements and we expect that this would hold true for many other scientific computing codes as well. 

\section{Acknowledgements}

The authors would like to thank Glenn Volkema for his assistance with this work throughout, many helpful discussions and also for providing useful feedback on this manuscript. GK would like to acknowledge research support from the National Science Foundation (NSF grant numbers: PHY-0831631, PHY-0902026), IBM, Sony and Nvidia. JM is grateful for support from the Massachusetts Space Grant Consortium. 

\section{References}

\begin{itemize}

    \item {\bf Websites:}

    [1] IBM's Cell site: http://www.research.ibm.com/cell

    [2] Sony PS3: http://www.us.playstation.com/

    [3] RoadRunner: http://www.lanl.gov/roadrunner/

    [4] Nvidia's CUDA: http://www.nvidia.com/cuda/

    [5] NSF's LIGO: http://www.ligo.caltech.edu/ 
        and several others (GEO, Virgo, TAMA, AIGO). 

    [6] ESA/NASA LISA: http://lisa.nasa.gov/

    [7] Cell Broadband Engine Architecture and its first implementation -- a performance view: 
        http://www.ibm.com/developerworks/power/library/pa-cellperf/

    [8] OpenCL: http://www.khronos.org/opencl/

    \item {\bf Papers:} 

    [9] L. Burko, G. Khanna, Accurate time-domain gravitational waveforms for extreme-mass-ratio binaries, {\it Europhysics Letters 78}, 2007, 60005; P. Sundararajan, G. Khanna, S. Hughes, Towards adiabatic waveforms for inspiral into Kerr black holes: I. A new model of the source for the time domain perturbation equation, {\it Phys. Rev. D 76}, 2007, 104005; P. Sundararajan, G. Khanna, S. Hughes, S. Drasco, Towards adiabatic waveforms for inspiral into Kerr black holes: II. Dynamical sources and generic orbits, {\it Phys. Rev. D 78}, 2008, 024022.

    [10] V. Agarwal, L.-K. Liu, D. Bader, Financial modelling on the cell broadband engine, {\it IEEE International Parallel \& Distributed Processing Symposium},  Miami, Florida, 2008.

    [11] M. Christen et al., Accelerating Stencil-Based Computations by Increased Temporal Locality on Modern Multi- and Many-Core Architectures, {\it New Frontiers in High-performance and Hardware-aware Computing (HipHaC)}, Lake Como, Italy, 2008.

    [12] Scherl et al., Fast GPU-Based CT Reconstruction using the Common Unified Device Architecture, {\it Nuclear Science Symposium / Medical Imaging Conference}, Honolulu, Hawaii, 2007.

    [13] J. Breitbart, G. Khanna, An exploration of CUDA and CBEA for a gravitational wave data-analysis application (Einstein@Home), accepted in {\it International Conference on Parallel Processing and Applied Mathematics}, Wroclaw, Poland, 2009.

    [14] S. Teukolsky, Perturbations of a rotating black hole, {\it Astrophys. J. 185}, 1973, 635.

    [15] W. Krivan, P. Laguna, P. Papadopoulos, and N. Andersson, Dynamics of perturbations of rotating black holes, {\it Phys. Rev. D 56}, 1997, 3395.

\end{itemize}

\end{document}